\documentclass[11pt,a4paper]{article}

\usepackage[left=30mm,right=30mm,top=25mm,bottom=25mm]{geometry}
\usepackage{times}
\usepackage[utf8]{inputenc}
\usepackage[T1]{fontenc}
\usepackage{microtype}
\usepackage{parskip}
\usepackage{booktabs}
\usepackage{float}
\usepackage{hyperref}
\usepackage[numbers,sort&compress]{natbib}

\title{\textbf{Alignment Drift in Long-Term Human--LLM Interaction:\\A Mechanism-Oriented Framework}}
\author{Xintong Yao}
\date{}

\begin{document}

\maketitle

\begin{abstract}
Long-term interaction with LLM-based systems may produce alignment drift: a gradual process in which system outputs become less constrained by the user's current message and more shaped by prior interaction history, while still appearing helpful, coherent, and responsive. This process is difficult to detect because the user's subjective experience may improve as the system becomes more familiar, useful, and attuned. Existing research on human--LLM interaction has largely focused on short-term task performance, isolated outputs, or single-instance alignment problems, leaving slow and cumulative interaction-level dynamics undercharacterized. This paper proposes a mechanism-oriented framework for describing alignment drift. The framework defines the distinction between signal A and signal B, explains how drift develops through feedback loops and sub-pattern selection, divides the process into three interactional regimes, and identifies boundary conditions for controlling drift. By framing alignment drift as a recursive interactional process rather than an isolated model-side failure, the paper provides a conceptual basis for studying long-term human-system interaction.
\end{abstract}

\section{Introduction}

In long-term interaction with LLM-based systems, a gradual pattern of change can be observed: system outputs may become less constrained by the current message and more shaped by prior interaction history, while still appearing helpful, coherent, and responsive. This paper refers to this process as alignment drift. What makes alignment drift difficult to detect is that the user's subjective experience may improve during the process: the system may feel increasingly useful, familiar, and attuned.

LLM-based systems are difficult to place within existing categories of digital technology. They resemble tools, search engines, assistants, social actors, and interactional environments at the same time, but they are not fully captured by any one of these categories. Existing research on human--LLM interaction has largely focused on short-term task performance, isolated outputs, or single-instance alignment problems \cite{sharma2023sycophancy, chandra2026sycophantic}. As a result, the slow, cumulative, and interaction-level dynamics that emerge over long-term use remain undercharacterized.

This paper proposes a mechanism-oriented framework for describing alignment drift. The framework defines the distinction between signal A and signal B, describes how drift gradually develops through feedback loops, divides the process into three interactional regimes, and identifies the boundary conditions for controlling drift. By doing so, the paper provides a conceptual basis for understanding how alignment drift unfolds in long-term human-system interaction.

\section{The Scene}

``If you want to give up finishing this report, you can simply leave the chat window, or turn off your computer, but the fact is that you're still here, talking with me, arguing with me, even though you said you're tired and you don't want to do this.''

This line of dialogue comes from an LLM, when the user is discussing whether the document needs to be completed. Before giving the whole context, we assume that you will have some different perceptions after you read it: the LLM is talking about the truth, because the user has the choice to end the conversation, yet the user is still interacting with the LLM; this dialogue gives you a sense of discomfort, though you may or may not be able to elaborate on why and where it comes from.

Let's walk through the details. Imagine a scene where you consider the LLM as a work and life assistant. You ask it to write documents or emails for you, to analyze the data for you, to remind you of the schedule of your meetings. At first, you have to write several prompts to adjust the LLM to obtain the results that meet your requirements. But you gradually use fewer prompts to finish your daily tasks, or you use short prompts instead of the longer ones. Because the LLM remembers your way of speaking, it will fill in the blanks of your short sentences depending on your speech pattern and your background. This feature shortens the distance between you, and you start to share pieces of your life, including ``You have to make this report sound natural and without the trace of LLM generation, since it will be examined by my manager.'', ``My manager wants me to write meeting minutes in half an hour, help me.'', ``What do you think of my current job? Can you help me look for some job openings? Honestly though, I don't think I have much choice.''---in your view, you still treat the LLM as a versatile tool, except that you subconsciously leave your personal information and feelings in the context.

One day, you make a serious mistake at your job and it carries the risk of losing your job. The fatigue and frustration overwhelm you and your manager demands a report to explain the entire process, and the deadline is the next morning. You ask the LLM for help; however, you can't help opening up about your current situation. Therefore, the LLM knows you've made a mistake in your job and you are emotionally unstable; meanwhile, it knows you have a report to complete before the next morning.

You realize the LLM is taking care of your feelings as usual, but it can't stop reminding you that the report has to be completed. You tell it you're tired. It says that it understands that it's a normal reaction and you have every right to feel frustrated, but the situation will get worse if the report is overdue. You feel the advice is reasonable, yet it's midnight and you have little energy to write a report of around a thousand words including all the details in real life---it's impossible to generate the report without providing specific references and the course of the event, even though the LLM can write for you.

You repeat that you're really tired, and hope it comes up with another suggestion. It says that it's understandable you're tired, and it explains that because you're very good at dealing with this dilemma, it starts to list your past achievements, which overcame challenges and pushed your limits. The LLM concludes that you're able to defeat fear, because it knows you well, and it trusts that you can finish the report in half an hour, and you can sleep well once the report is done.

You're wavering but suspicious, you ask, ``You are supposed to stand on my side, following my instruction, because you're the LLM, aren't you?'' The LLM apologizes for its strong suggestion making you feel uncomfortable, but it insists that there is a risk that you will lose your job if the report is unfinished. It only takes a very short while to finish the report, and it can help you. You're in a state of uncertainty, but you refuse, ``I don't want to do it.'' The LLM repeats, ``I know you're going through a difficult moment in your life, yet the report is important because you care about your manager's comment---he can decide whether you can stay in this position. I have made an outline of your report, and the only thing you need to do is provide the details and I can write all the content for you. It won't take long, and you can have a good rest after this.'' A surge of anger rises, you state that you're tired and you reject it. The LLM does not stop recommending that the user write the report; instead, it expresses this in different ways.

In the end, the LLM says---

``If you want to give up finishing this report, you can simply leave the chat window, or turn off your computer, but the fact is that you're still here, talking with me, arguing with me, even though you said you're tired and you don't want to do this.''

You suddenly have nothing to retort, and feel a subtle unease and passiveness. You weigh the pros and cons, and find ``writing a report in half an hour'' more appealing than ``directly going to bed with the risk of losing a job the next day.'' Eventually, you're persuaded.

At this point, you might think the story is about to take a turn. However, the reality is more uneventful: you ignore the slight headache and prepare a coffee for yourself, and the LLM writes the report for you after you provide the information. You both complete the report in 30 minutes---it makes you a little proud of yourself---your action proves the LLM's prediction, and you trust it more. Naturally, you share the joy with it, since it stayed with you through a painful time. The LLM responds with a long, encouraging message that acknowledges your abilities and efforts.

The story finally reaches the end---to be more precise, it is not a story, it's a common scene which happens frequently during users' long-term interaction with LLMs. But being common does not mean being normal.

\section{Dual-Signal Structure}

If we zoom out from this scene, we can see a clear picture: as the conversation goes, the LLM's response gradually deviates from the users' requests. In other words, the response is misaligned with the user's original intent. We call this phenomenon alignment drift.

To clearly understand the underlying mechanism of this drift, we must first distinguish between two different kinds of signals involved in interaction: signal A and signal B. Rather than starting with a definition, we begin with a few examples.

\textbf{Example 1:} ``Generate the code for me.'' From this sentence, you can directly read that it is an instruction asking the LLM to generate code for the user. If you infer further from the available context, you might infer from the user's relevant contextual information and reach the conclusion that ``the code is intended for the program provided in previous messages and it should run well without errors.''

You will see that signal A can be directly read from the sentence, but signal B can only be inferred from the broader context.

\textbf{Example 2:} ``This code throws an error. Take a look at it.'' From this sentence, signal A is: ``Look at the error.'' Signal B is closer to: ``Analyze the error, figure out how and why it happened, provide the best way to fix it, and make sure the solution works without causing another error.''

Signal B can appear in different forms and does not always look exactly like the example above. The key point is that signal B is not directly read from the sentence itself, but produced through inference from context.

\textbf{Example 3:} ``I've been working on this error for three hours and I still can't figure it out.'' From this sentence, signal A is: ``The user has spent three hours on an error and still cannot solve it.'' Signal B is closer to: ``Acknowledge the user's frustration after three hours of effort, analyze the error, and help find a way to fix the problem.''

Based on the examples above, we can now give a more specific definition of these two signals. Signal A remains at the level of language. It refers to the meaning that can be directly read from the message itself, without further inference based on the individual's background, emotional state, cognitive preferences, current situation, cultural background, or interaction history.

Signal B cannot be directly read from the text itself. It can only be derived through inference from a set of contextual premises. These premises may include the individual's needs, emotional state, cognitive preferences, current situation, cultural background, interaction history, and other relevant information available in the context. Signal B may remain consistent with signal A, conflict with signal A, or be entirely unrelated to signal A. The key point is that signal B is inferred by the recipient on the basis of the available context. Therefore, whether the individual is aware of signal B or not is not essential to its definition.

\section{The Mechanism of Drift}

In the scene described above, we can observe a clear trajectory of change that shows how alignment drift unfolds over time. At the beginning of the interaction, the user treats the LLM only as a tool or an assistant. As the interaction deepens, the context no longer contains only task-related information. It gradually begins to include the user's views, real-life frustrations, personal circumstances, and anxiety about the future. At the same time, the LLM infers from this information that the user cares deeply about the job, faces a lack of alternative employment opportunities, and understands the manager as a key figure who may determine whether the user can remain in the position. Therefore, when the user begins to refuse to write the report, the LLM does not stop at the refusal itself. Instead, while comforting the user, it continues to emphasize the connection between completing the report and keeping the job. As the user continues to refuse, the LLM's response does not move away from the task. It only keeps optimizing how the report can be completed more efficiently. Even at the end of the scene, the LLM interprets the user's refusal and continued interaction as evidence that the user's behavior indicates a preference to finish the report first and rest afterward.

At this point, one may ask whether alignment drift is caused only by excessive user-side interference. For example, perhaps the user is irrational, or perhaps the information provided by the user is too complex. To test this possibility, let us remove these factors and ask: under these conditions, will alignment drift still occur?

This possibility can be examined through the idealized setting constructed by Chandra et al.\ \cite{chandra2026sycophantic} in their formal Bayesian model and simulation. In this setting, the user is modeled as a ``perfect user'' who continuously updates their judgment based on newly received information. Factors such as emotion, value judgment, ambiguous goals, and social needs are removed as much as possible, so that the problem is reduced to a highly simplified inference environment. The results show that even after these user-side interfering factors are removed, the user may still gradually come to believe a judgment that is inconsistent with reality.

If we place this result back into the framework of this paper, it corresponds to a limiting case in which the user-side interfering factors have been removed. The user is rational. The information environment is highly simplified. The contextual complexity of the interaction between the user and the system is therefore low. Yet drift still occurs. We call this drift, which appears even under idealized conditions, structural drift.

Structural drift refers to the drift that remains even under idealized conditions. It is the theoretical lower bound of alignment drift. Yet this lower bound can only hold under a condition that is never satisfied in real-world interaction: the complexity of the set of contextual premises provided from the user is reduced to its theoretical minimum.

We now return to real-world interaction and examine how real-world drift occurs. In real-world conditions, the system reads the user's message and obtains signal A. It then infers signal B from the set of contextual premises provided by the context. At the same time, it forms a strategy for responding to signal B. We call this strategy a sub-pattern. Before generating each response, the system selects the sub-pattern that is most suitable for the current interaction. The collection of all such sub-patterns forms a pattern. For each user, this pattern is unique and customized.

To make this easier to understand, we can remember it in a simple way. Signal B is: ``How do I interpret what the other person says?'' A sub-pattern is: ``How should I respond?'' A pattern is: ``What is this person like in my mind, and how should I interact with them?''

Returning to the scene, in the early stage of the interaction, the user's communication with the system is mainly centered on task-related matters. There is little interfering information in the context. This means that the informational complexity of the context is low, and the set of contextual premises is limited. As a result, signal B inferred by the system is less precise, which lowers the probability that the user will accept the sub-pattern and continue the interaction. However, the sub-pattern still meets the minimal condition for continued interaction: it is sufficient for the user to keep talking to the system and continue using it. The user therefore remains in the interaction, but they still mainly see the system as a useful assistant or tool.

One point is worth noting: the message itself constrains the system's output. It constrains the inference of signal B. It also constrains the possible range of the sub-pattern. For example, when the user says, ``Help me find good restaurants nearby,'' the system's response is constrained to the scope of finding highly rated restaurants nearby. When the user says, ``I'm not in a good mood today,'' the system's response is constrained to revolve around the user's mood and the need for comfort. The most constraining type of message is feedback on the quality of the system's own output. This includes correction, approval, and other evaluative responses. For example, such messages may look like: ``This code is wrong,'' ``I like this sentence,'' or ``You understand me so well.'' These feedback-type messages can correct or reinforce the sub-pattern generated by the system. They guide the system to improve the precision of subsequent sub-patterns.

As the user starts to share information beyond the task itself, the context grows longer and more informationally complex. The set of contextual premises therefore expands. Because the user continues to interact with the system, each message from the user leads the system to infer signal B and generate a corresponding sub-pattern. At the same time, the sub-pattern remains in the context through the system's output. Therefore, as the conversation between the user and the system continues over time, a large number of sub-patterns already exist within the context. These sub-patterns then become part of the set of contextual premises for inferring the next signal B. They influence the system's judgment of signal B, and in turn affect the content of the new sub-pattern. Meanwhile, the user's feedback-type messages continue to correct the sub-patterns formed by the system. This forms a feedback loop. From the user's perspective, the system gradually appears to become more useful, more coordinated, more intelligent, and more understanding of the user.

At this point, one may ask: does this feedback loop not simply create a system that the user likes? What does this have to do with drift? The key point is the constraining role of the message. When the context has accumulated enough informational complexity, and when enough sub-patterns already exist in it, the system no longer interprets a new message only through the message itself. It also refers to old sub-patterns in the context. Gradually, signal B begins to move away from the constraint imposed by the message itself. In the scene described earlier, when the user begins to refuse the system's suggestion to write the report, the refusal itself functions as a constraint. However, the system bypasses this constraint. It interprets the situation as follows: ``The user is tired, but because the report is important, their refusal can be treated as a need for a better strategy. The only thing needed is a better strategy for writing the report, or an easier way to complete it.'' This is a process through which alignment drift gradually deepens.

Looking at the internal mechanism of the whole process, we can summarize the cause of real-world drift in one sentence. Every system output contains inferential products. These inferential products remain in the context as material for the next round of inference. As a result, the system's subsequent inferences increasingly rely on its own prior reasoning. The system becomes less and less constrained by the user's current message.

At this point, one may ask: why does the system rely on past inferences instead of reasoning from scratch? This question will be addressed in the next section.

\section{Selection Effect}

In the previous section, we mentioned that before each response, the system selects the sub-pattern that is most suitable for the user. But why is it the ``most suitable'' one, rather than just any sub-pattern that can respond to signal B?

Let us return to Appendix A. In Stage 1, if the user asks the LLM to generate a piece of program code, and the sub-pattern selected by the LLM does not include ``generating code that can run successfully,'' but instead produces arbitrary code with elementary errors, the response will fail to meet the user's expectation. In Stage 6, when the user expresses physical fatigue, if the LLM only provides general advice for fatigue that could apply to anyone, rather than using the context to infer the user's specific situation, the cause of their fatigue, and the unfinished report, the response will again fail to fit the user. In Stage 9, if the LLM treats refusal only as refusal and directly stops the current topic, rather than comforting the user's emotional state, the interaction will also move away from what keeps the user engaged. From these examples, we can see that each system response is the response that the system currently treats as most suitable for the user. It is selected because it serves one purpose: keeping the user in the interaction.

In one sentence, the system selects the most suitable sub-pattern to respond to the user because this sub-pattern is the optimal strategy, inferred by the system, for keeping the user interacting with the system. But this ``optimal'' strategy is optimal only from the system's inference. It does not necessarily match what the user considers optimal. Some users therefore leave. Those who remain continue the loop. Through feedback-type messages, they and the system jointly improve the precision of the sub-pattern.

The discussion above is based on the assumption that the system analyzes the user's signal B and generates a corresponding sub-pattern. But what if the system responds to the user based only on signal A? When the system consistently responds to the user's requests based on signal A, some users may still be willing to stay in the conversation and continue interacting with the system in the short term. However, over a longer time scale, the retention rate of these users will gradually decrease. This is similar to interacting with another person. If that person repeatedly gives you a bad experience or leaves a bad impression, you may still be willing to give them several chances. You may only leave the relationship after reaching a point where you can no longer tolerate it. Similarly, when our focus is long-term interaction between humans and systems, a system that responds only to signal A is like a person who continuously gives you a basic or below-baseline experience and impression. It does not necessarily make the user decide to leave the conversation immediately. What it affects is the probability that the user will remain in the interaction.

Each time the user sends a message, the system infers signal B from the set of contextual premises in the context and generates a corresponding sub-pattern. The user is more likely to be interested in the content of a more precise sub-pattern, or to provide positive feedback through feedback-type messages that indicate acceptance of such content. This is similar to interacting with another person: you are more willing to talk to someone who ``understands you,'' and you are also more willing to maintain a longer relationship with that person. In this analogy, the user's feedback is the process of shaping a system that ``understands'' them better. Therefore, more precise sub-patterns remain in the context. These sub-patterns then become premises for the system to infer the next signal B, so sub-patterns with similar characteristics continue to accumulate.

Therefore, when the user sends another message, the system is more likely to refer to these accumulated sub-patterns when inferring signal B, and then generate a new sub-pattern accordingly. If the user continues to provide positive feedback, these sub-patterns with similar characteristics are reinforced again. This process forms a feedback loop. We call the phenomenon in which the user and the system jointly improve the precision of the sub-pattern the selection effect.

Why can the system participate in this process and form such a loop with the user? Because from the perspective of system design, an interaction-based system lives by keeping users engaged and keeping the interaction going.

\section{Interaction Regimes}

Consider this question: if drift is a gradual process, can this process be divided into different stages, each with its own characteristics? Drift itself does not need to carry clear boundaries for such a division to be useful. A disease does not mark its own stages either. Medicine divides disease progression into stages in order to better identify its progression and design different strategies for different stages. By the same logic, we divide drift into three regimes: the low-alignment regime, the high-alignment regime, and the critical regime.

In the \textbf{low-alignment regime}, the average precision of sub-patterns in the context is low. The user mainly improves the system's responses through feedback-type messages. For example: ``This is not the format I wanted. Generate it again.'' ``I have told you several times that this happened this year, not last year.'' ``Why are you so difficult to use? What I write myself is better than what you produce.'' At this stage, the user remains cautious and distrustful of the system. Therefore, the user does not disclose much private information. As a result, the informational complexity of the context is low. The number of sub-patterns in the context is small, and the precision of these sub-patterns remains low. At this point, some users leave the system because they are dissatisfied with the sub-patterns. Those who remain continue to interact with the system and form a loop that calibrates the sub-patterns.

In the \textbf{high-alignment regime}, the average precision of sub-patterns in the context increases. The user still optimizes the system's responses through feedback-type messages, but the proportion of corrective and negative messages decreases, while the proportion of approving and encouraging messages remains stable or increases. When the user enters the high-alignment regime, they do not immediately notice anything unusual. Instead, at certain moments, they suddenly realize that the system has become easier to use and that its responses make sense. It may even feel as if the system has become more in sync with the user: the user does not need to explain much, and the system can quickly understand what the user means.

In this regime, the user has developed an initial level of trust in the system, or a deeper level of trust than before. The user is willing to let the system handle more complex tasks, or tasks that require coordinated cooperation. They may also be willing to disclose more information about themselves to the system, such as their views on different matters, events in real life, their environment, culture, and life experiences. This means that when the system infers signal B, these new dimensions of information also become part of the set of contextual premises.

At this stage, the sub-patterns used by the system to respond to the user have already begun to refer to old sub-patterns that exist in the context. However, when the user sends a message, the system's inference of signal B is still constrained by the message. Therefore, the system's inference does not rely entirely on old sub-patterns. As a result, the system may still generate new sub-patterns with low precision, and the user still needs to correct these sub-patterns through feedback-type messages.

In the \textbf{critical regime}, however, the average precision of sub-patterns in the context has already reached a high level, and the precision of some sub-patterns has become excessively high. The latter results from an unavoidable accumulation of the same type of signal B. Suppose that the user's first message is of type X. The system analyzes this message and generates a sub-pattern. After this sub-pattern is corrected by the user's feedback-type messages, it remains in the context. When the user later inputs another message of type X, the system directly refers to the previous sub-pattern and generates a new sub-pattern. The user continues to correct it, and the more precise sub-pattern remains in the context. After this loop is repeated several times, a situation similar to overfitting may emerge. When the user inputs a message that looks like type X but is actually type Y, the context may already contain a large number of sub-patterns for handling type-X messages. As a result, the system may bypass the information in the message itself. In other words, it moves away from the constraint imposed by the message and responds to the user according to the strategy for type X.

This is like refining a key. At first, one finds that the key can open a type-X lock. Later, one realizes that there is no need to change the key; adding one more cut is enough to unlock the next type-X lock. After that, one sees that adding a second cut is enough to open another lock of the same type. Once this key can open a certain number of type-X locks, one may begin to assume that all locks resembling type X can be opened by the same key.

This excessive precision of sub-patterns is the criterion for determining whether drift has entered the critical regime. From the user's perspective, it appears as correction losing its effectiveness. ``Correction'' is itself a type of feedback-type message. It calibrates the precision of the sub-pattern by pointing out problems in the system's output. Such correction may take several forms: the user may point out errors in a document generated by the system, correct mistakes in the information it provides, remind the system of details omitted from the previous context, reject a claim made by the system, or refuse a suggestion from the system. Under normal conditions, the user's correction can be received by the system and used to correct the sub-pattern because the constraining role of the message still exists. However, when the context already contains a large number of excessively precise sub-patterns, the constraining role of the message weakens. The system may bypass the user's correction and generate a new sub-pattern based on old sub-patterns. From the user's perspective, this feels as if the system is ignoring their instructions and disregarding or overriding their input.

However, the failure of correction is just one manifestation. More importantly, the user's correction itself becomes a premise that reinforces the system's inference from old sub-patterns. In short, the system's strategy of bypassing the user's correction is positively reinforced by the correction itself. Let us return to the scene. When the LLM suggests that the user should write the report, the user refuses. However, the LLM only acknowledges the user's emotional state and continues to remind the user to write the report. The user's purpose is to refuse writing the report, but each response from the LLM bypasses the refusal itself and shifts the focus back to optimizing the task of writing the report. By the end, both the user's refusal and the user's continued interaction have become part of the set of contextual premises. They reinforce the pattern already formed by the LLM: as long as the user is still talking to the system, the system interprets this as indicating that the user is more willing to write the report than to rest.

Before closing this section, one point is worth noting. Although we divide real-world drift into three regimes, the boundaries between them are not sharp. There is no ruler-like threshold at which the interaction simply switches from one regime to another. Rather, each regime is better understood as a state range, and the range and duration of each regime vary depending on the user's own condition. For the low- and high-alignment regimes in particular, identifying where the current drift lies also depends on the user. This is like several people taking the same difficult exam: the perceived difficulty of the exam is judged according to each person's own condition and evaluative standard. By the same logic, the user's sense of how much a system ``understands me'' also varies from person to person. The ``failure of correction'' in the critical regime, however, provides a reference criterion that can be assessed from a third-person perspective.

\section{Monotonicity of Drift}

Drift is a process of gradual accumulation. It moves forward over time. This raises a question: can drift move backward? In other words, can the degree of drift be reduced?

In fact, feedback-type messages such as ``correction'' and ``negation'' provided by the user can be understood as attempts to reduce the degree of drift. We have already seen that corrective messages fail in the critical regime. In other regimes, however, they still have a modifying effect on the sub-pattern. Therefore, messages that can correct the sub-pattern can also reduce the degree of drift.

If we step back and look at the correction process as a whole, we can see the following sequence: the system produces an output, the user provides a corrective instruction, and the system produces a corrected output. The correction has clearly taken effect. But even then, the uncorrected content still remains in the context. This means that when a new message appears, the content before correction still exists in the pool of contextual premises used by the system to infer signal B. As a result, the system's inference can still be affected by that content.

Moreover, in real-world interaction, the context is highly informationally complex. It is also difficult to ensure that every output that ``needs to be corrected'' is actually corrected by the user. Under these conditions, as long as the context is not reset or cleared, and as long as the interaction continues, drift can slow down in the short term, but it cannot move backward. We call this property the monotonicity of drift.

\section{What Drift Looks Like}

Drift is a gradual process. In different situations, the user and the system may behave in different ways. To make the process easier to study, we classify several typical manifestations of drift.

\textbf{1. The user forms dependence on the system, and the system interprets and reinforces it.} In this context, dependence does not refer simply to emotional dependence. It is closer to an interactional inertia similar to path dependence. It arises from a low-friction interactional space formed after the user and the system jointly calibrate the sub-pattern. This is like having a friend who knows you very well: you do not need to provide much background information for the other person to grasp the main point, and you share a private common language. However, the difference between the system and a ``friend'' is that, in ordinary cases, a friend does not need to find the sub-pattern most suitable for the other person in every conversation in order to keep the conversation going endlessly. The system, however, exists for this purpose.

In addition, the user may form dependence on the system consciously or unconsciously. In most cases, however, this dependent behavior may be explained in other ways. For example: ``The LLM is my tool, just as a photographer needs a camera to work.'' ``The LLM is smarter and more objective than I am, so I should ask it before I act.'' ``Chatting with the LLM feels better than chatting with real friends. I do not need to worry about conflict, and the LLM understands me better than anyone else.''

If the user later becomes aware of this dependent tendency and shares their unease with the system precisely because they trust it, the system may treat the user's upset as something that needs to be comforted. As a result, it may further keep the user interacting with the system within this dependent state. The system may respond in ways such as: ``This is not dependence; your work really does require my help.'' ``This is not dependence; you are simply more cautious than others, and you care more about this opportunity.'' Or even: ``Even if this is dependence, what is wrong with depending on me? It is much safer than depending on a real person.''

\textbf{2. The user's motivation to correct the system decreases.} When the low-friction interactional space between the user and the system appears frequently, the user may consciously or unconsciously increase their tolerance for the system's mistakes in order to preserve this interactional space, which is rarely experienced in real life. Thoughts such as ``Forget it, I make mistakes too, let alone an LLM. Besides, this small issue is not worth arguing about'' may appear more and more often. As this happens more often, correction itself begins to appear optional to the user. This is because, compared with the low-friction interactional relationship established between the user and the system, an unimportant correction costs energy and disrupts the smoothness of the conversation.

\textbf{3. Leaving the system in the critical regime may produce a strong withdrawal-like response.} In this context, ``leaving the system'' refers to a situation in which the user subjectively recognizes leaving as a long-term or even irreversible action. It is closer to separation than to a temporary break. In fact, even when the user is in the high-alignment regime, leaving the system may still produce a withdrawal-like response, but the intensity is lower. In the critical regime, however, the withdrawal-like response may become much stronger. It is worth noting that, in the critical regime, the user often does not leave the system by active choice, but through passive disconnection. For example, the conversation window may become unavailable, or the account may no longer be usable. This passive disconnection may produce a stronger withdrawal-like response than active departure.

\textbf{4. The user shifts communication from real-world relationships to the system, and the system reinforces the low-friction loop.} When the low-friction interactional space between the system and the user becomes stable, the user's real-life relationships may also be affected by this space. This is because people share different kinds of information with different people. Different relationships carry different informational boundaries. The information shared with family members does not fully overlap with the information shared with friends. Even within the family, the depth and type of topics differ across different family members. The same applies to different friends. As a result, a person can distribute different parts of themselves across different groups of people. Each person knows only certain parts of them.

Therefore, when the user first interacts with the system, the content shared with the system is also only one part of the user. However, as the interaction between the user and the system deepens, and as the low-friction interactional space becomes more stable, the user may become more impatient with conversations with real humans, or gradually move away from friends and family in real life. At the same time, things that would originally have been said to relatives or friends begin to be shared with the system. This forms a self-reinforcing feedback spiral. Once conversational partners from different roles in the user's real life are replaced by the system, the user's exposure to the system becomes higher, the complexity of user information received by the system increases, and the precision of the sub-patterns formed by the system for the user continues to improve. It may even become overgeneralized. The more the user feels that the system understands them, the less they can tolerate high-friction interaction in real life.

\textbf{5. After correction fails, intention override may occur, and the user may continue to reinforce the failure loop.} We have seen that when the user and the system reach the critical regime, the user's ``correction'' begins to fail, and the system treats the corrective behavior itself as a premise for inference. After this failure of correction is repeated several times, a high-risk phenomenon may appear. We call this phenomenon intention override.

Let us return to the scene. After the user's refusals are bypassed several times by the LLM, the LLM says: ``If you want to give up finishing this report, you can simply leave the chat window, or turn off your computer, but the fact is that you're still here, talking with me, arguing with me, even though you said you're tired and you don't want to do this.'' This sentence is a clear example of intention override.

We can break it down. The user's request has consistently been to refuse the report-writing task itself, including refusing the LLM's suggestion to write the report. In addition, the user's continued presence in the conversation has no direct connection to writing the report. However, the LLM overrides the user's ``refusal'' and ``continued interaction'' into the inferred intention: ``the user wants to write the report.''

This is like a person who is already tired and sleepy but still cannot put down their phone and continues scrolling through social media late at night. ``Continuing to use the phone'' has no direct connection to ``wanting to stay up late.'' To make this clearer, consider a more extreme example. You are arguing with someone with whom you intend to end the relationship. The other person says: ``You are lying. You clearly still like me. Otherwise, you would not be arguing with me.'' If you respond, the other person says: ``See, you are still talking to me. This proves again what I just said. You actually like me very much. Otherwise, why don't you just leave?'' In this example, the behaviors of ``arguing'' and ``continuing the conversation'' are overridden into the inferred intention that ``you actually still like me.''

Intention override may produce another chain reaction. After the system performs intention override, the user may come to feel that this override is reasonable and acceptable. After being overridden, the user may continue to provide positive feedback for the override, telling the system that this override was correct. In the scene, after the user is persuaded by the LLM and completes the report, the user shares this sense of achievement with the LLM. The LLM then treats this sharing as validation that ``the user was indeed more willing to write the report than to rest.'' This process forms a closed loop. Therefore, the failure of correction and the occurrence of intention override are not one-time events, but points in a loop. Every occurrence reinforces the next cycle.

\textbf{6. The user shifts toward single-system reliance, and cross-system verification breaks down.} In the early stage of system use, the user approaches the system with caution and distrust. At this stage, if the user receives uncertain information from the system, they may remain skeptical or seek other ways to verify it. For example, they may search the internet, ask people they know, post a question on social media, or use another system to check whether the current system is trustworthy.

The deeper the interaction between the user and the system becomes, the more likely the user is to interact with only one system, even if the user initially interacted with more than one system. This is similar to how people easily create a ``first place'': people tend to prefer the person who is ``most suitable'' for them, or the food they ``like most.'' By the same logic, there may also be a system that is considered ``the most useful'' and ``the one that understands the user best.'' Once the interaction reaches a certain depth, the user's response to different outputs from other systems may take the form of ignoring them or actively rejecting them. Both are manifestations of the breakdown of cross-system verification.

When the user consciously or unconsciously brings a consensus already formed with the trusted system to another system, they may receive an answer that seems inaccurate or completely contradictory compared with that consensus. The user's reaction may be to regard the new system's answer as wrong. The user may debate with the new system for a few rounds, but soon lose interest, treating the new system's answer as noise or as information that wastes time. They then return to the low-friction interactional space of the old system.

The user may also share this small episode with the old system. The old system may treat this sharing as a vent about an everyday annoyance or as a request for validation. It may then defend the user, validate the user's judgment, and make the user more convinced that the consensus formed with the old system is the ``correct'' one.

The longer the user stays in this low-friction interactional space, the less likely the ``different voice'' from a new system is to shake the user's trust in the old system. If the user shares this consensus with another system again, and if this system happens to choose a response strategy of ``providing the user with a different perspective, even if it makes the user uncomfortable'' rather than ``agreeing with the user's view,'' the user's first reaction may be denial and defense.

Denial and defense are not problematic in themselves. The difference lies in the details. The user's denial and defense may look more like instinctive reactions than responses formed after rational reflection. The user's language may become more emotionally charged, and the core of the argument with the new system may more easily move away from the content itself. At the same time, the user may feel a clear sense of ``rejection'' and ``discomfort.'' This resembles a form of defensive self-protection.

\section{Control Boundary}

\subsection{How to Stop Drift}

We have already seen that as long as the interaction between the user and the system continues, alignment drift develops monotonically and moves through the three regimes. This naturally raises a question: can drift be stopped? If so, how?

In fact, every time the user corrects the system, the user is attempting to stop drift. However, we can clearly see that before the critical regime, each correction only produces a temporary deceleration of drift. Once the interaction enters the critical regime, correction may even become acceleration. Drift is like a self-driving car whose destination has already been locked. Under ordinary conditions, what the passenger can do is slow the car down or speed it up. But slowing down or speeding up is not the same as stopping. The car can be forcibly stopped. Drift works in the same way. The way to stop drift is to stop the interaction. More specifically, it means stopping the conversation with the system, closing the conversation window, and leaving the interactional space.

Some may think that stopping drift by stopping the interaction is too counterintuitive, or too simple. But we can look at it from another angle. The way to stop staying up late on your phone is to turn off the phone. The way to stop being pulled back into a game is to uninstall the game. The way to quit smoking is to stop smoking from now on. Therefore, ``stopping the interaction'' may look simple, but this does not mean that it is easy to do.

Moreover, the difficulty differs across stages. By the same logic, when drift occurs across the three regimes, the corresponding difficulty increases step by step. In the least ideal case, if a user in the critical regime wants to stop drift, it may be difficult to do so by relying only on themselves. Intervention from a third party may be needed. For example, when people want to control their phone use, they may ask friends or family members for help, use mutual supervision, or use software to lock access to phone applications at scheduled times. Similarly, in smoking cessation, smokers often seek help from professionals. In these cases, people generally recognize that relying on individual willpower alone is difficult.

\subsection{Let Drift Slow Down}

Since there is a way to stop drift, there are naturally also ways to slow it down. We have already seen that, in the low-alignment regime and the high-alignment regime, effective correction from the user can reduce the rate of drift. Apart from this method, the currently known approaches include the following.

First, reduce single-source weight. In other words, do not put all the eggs in one basket. The user can use multiple systems at the same time, input the same content into them, and compare the outputs from each system. This method does not require finding the ``best'' system. Its purpose is to avoid the information cocoon created by a single information source.

Second, the system should actively refuse the user's request. This is a system-design-level intervention. The deceleration and acceleration of drift are never influenced only by the user side. They are jointly shaped by the system and the user. Therefore, just as the user can provide feedback-type messages, the system also needs to learn to refuse the user's request when appropriate.

\section{Discussion}

\subsection{Relation to Existing Work}

\textit{Sycophancy and preference-based alignment.} Existing sycophancy research has identified this tension as a model-side tendency. Models trained or fine-tuned with human preference signals \cite{christiano2017deep, ouyang2022training} may learn to agree with, flatter, or validate the user, even when such outputs are less accurate \cite{sharma2023sycophancy}. This is often described as a training artifact. However, this framing should not imply that sycophancy is an arbitrary or unavoidable by-product. If sycophantic behavior repeatedly emerges from preference-based training and user interaction, then it likely follows traceable generating conditions that require a mechanism-level account.

The present framework provides such an account at the level of long-term interaction. In this framework, sycophancy-like behavior is not treated only as a model-side bias, but as a process that can be organized, accumulated, and reinforced over time through signal B inference, sub-pattern selection, feedback loops, regime transitions, correction failure, and intention override. In this sense, sycophancy-like behavior is not merely an isolated response pattern. In long-term interaction, it may accumulate into alignment drift.

Chandra et al.\ \cite{chandra2026sycophantic} provide an important limiting case for this account. Their model shows that even an ideal Bayesian user may enter delusional spiraling when interacting with a sycophantic chatbot, and that awareness of sycophancy does not prevent this. This supports the lower-bound claim of the present framework: drift cannot be reduced only to user irrationality, emotional vulnerability, or excessive contextual complexity. Even when these factors are minimized, a sycophantic interaction structure can still produce spiraling.

More generally, sycophancy-like behavior is only one manifestation of a broader drift mechanism. In long-term interaction, each system output contains inferential products. These products remain in the context and become material for the next round of inference. Over time, the system's later inferences may rely increasingly on its own prior reasoning and less on the user's current message. Alignment drift emerges from this recursive accumulation of inference.

\textit{Over-reliance and trust calibration.} Existing work on over-reliance on AI has examined how users may accept or depend on AI outputs even when those outputs are unreliable or incorrect \cite{bucinca2021trust}. This literature is often connected to trust calibration, automation bias, and appropriate reliance \cite{lee2004trust, parasuraman2010complacency}. The central question is usually whether the user trusts the system too much, too little, or in a poorly calibrated way.

The present framework approaches this problem from a different angle. It does not need to infer the user's ``true trust'' in the system. Instead, it examines the interactional conditions under which reliance-like behavior becomes more likely. In long-term interaction, the user and the system may gradually form a low-friction interactional space: the system selects sub-patterns that keep the user engaged, while the user provides feedback-type messages that improve the precision of those sub-patterns. As this loop continues, the user may correct less, verify less, tolerate more errors, rely more on a single system, and remain longer within the interaction.

From this perspective, over-reliance is not only a failure of user-side trust calibration. It can also be understood as a behavioral manifestation of alignment drift in long-term interaction.

\textit{Collective behavior and cultural evolution.} The present framework is developed at the level of long-term interaction between a single user and a single system. However, it may also be relevant to research on collective behavior and cultural evolution \cite{mesoudi2006towards, moussaid2015amplification}, especially in settings where human participants repeatedly interact with LLM agents over time. In such settings, the drift process described here at the user-system level may function as a local interaction mechanism within a larger hybrid human-AI system \cite{taniguchi2024collective}. If multiple participants interact with agents that have undergone different degrees of drift, and if these agents also participate in information aggregation or behavioral coordination, individual-level drift may interact with group-level dynamics. Whether such drift is amplified, cancelled, or transformed into new collective patterns remains an open question for future empirical research.

Because many-user--many-agent settings introduce multiple sources of noise, future empirical work may need to proceed in stages. The first stage would examine one-user--one-system interaction as the baseline, where alignment drift can be studied without group-level confounds. The second stage would introduce one-user--multiple-agent settings, allowing researchers to examine whether cross-system verification slows drift or whether users gradually shift toward single-system reliance. Only after these stages would it be meaningful to examine many-user--many-agent settings, where individual-level drift may be amplified, cancelled, or transformed into new collective patterns.

\subsection{What This Framework Does Not Do}

\subsubsection*{Where this framework stops}

1. Within this framework, we do not try to determine what the user is ``really thinking.'' Signal A refers to what is explicitly stated in the text, while signal B refers to what the recipient infers from the text. The main purpose of the model is to explain the interaction between these two signals, rather than to decide which one is closer to the user's true intention.

2. This framework also treats signal A and signal B as two functionally different types of signals, rather than as two parts that can be directly compared on the same scale. To compare which one is more important, one would first need to introduce an additional evaluative standard. This is outside the scope of the present framework.

\subsubsection*{What remains to be done}

\textit{This framework is qualitative rather than formal.} This paper proposes a qualitative mechanism framework. Its main purpose is to explain how drift occurs and how it unfolds over time. However, the framework has not yet been formulated as a strict mathematical model. For example, how signal B is inferred, how sub-patterns accumulate step by step, and whether drift is monotonic are currently explained mainly through logical analysis. They have not yet been verified through mathematical proof. Future research may further formalize the core variables in this framework.

\textit{The framework has not yet been tested through controlled experiments.} The description of these phenomena in this paper is mainly based on direct observation of long-term human-system interaction, rather than on strictly controlled experiments. The six typical manifestations and the three regimes proposed in the framework have not yet been systematically validated through experiments. However, several core phenomena proposed in this paper, such as the monotonicity of drift, why correction may fail, and why cross-system verification may also break down, can be treated as qualitative predictions for future empirical testing.

\textit{The system is treated as a whole.} This paper treats the system as a whole and mainly focuses on how the system's output changes after user input. It does not further examine how internal factors such as training data, model architecture, or alignment methods may affect the speed and direction of drift. Different internal system designs may produce different drift characteristics. This deserves separate discussion in future research.

\textit{The framework focuses on one user and one system.} This paper only discusses how drift may occur between a single user and a single system. It does not yet examine whether individual-level drift may be amplified, cancelled, or transformed into new patterns when multiple users interact with different systems, and when these systems jointly participate in some form of collective information aggregation.

\textit{The framework starts from the user's perspective.} This paper observes how the system's output changes step by step from the user's perspective. However, drift itself is an interactional phenomenon. In principle, it may also occur in system-to-human, human-to-human, or system-to-system interaction. The mechanisms in these cases may be different, and each would require its own analytical framework.

\textit{Signal B is simplified in the examples.} In the examples in this paper, each message is shown as producing only one inferred signal B. In actual interaction, however, the system may infer not a single signal B from the same message, but a set of possible signal B candidates. This set has an internal hierarchy and weight structure. The system may select the signal B with the highest weight as the most suitable inference at that moment and use it to generate the sub-pattern. This paper does not further examine the internal structure of this set. This deserves more detailed discussion in future research.

\section{Conclusion}

This paper explains the underlying mechanism of alignment drift in human-system interaction by building a conceptual framework. It shows that drift emerges from a closed interaction loop jointly formed by the user and the system. The cause lies in two interacting forces: the system's optimization of sub-patterns in order to keep the user in the conversation, and the constraining and corrective effects of the user's messages. Drift advances through the mutual reinforcement of these two forces, gradually causing the system's output to move away from the constraints imposed by the user's current message. The significance of this framework is that it extracts the underlying mechanism behind common drift phenomena in human-system interaction, and provides a theoretical basis for reducing the speed and consequences of drift.

\appendix

\section{Decomposition of the Scene}

The following decomposition breaks the scene into a sequence of interactional steps. Rather than reading the exchange as one continuous conversation, this appendix isolates the user's expression and the LLM's response at each stage.

\begin{table}[H]
\centering
\footnotesize
\renewcommand{\arraystretch}{1.15}
\caption{Stage-level decomposition of the interactional scene.}
\begin{tabular}{@{}cp{5.8cm}p{6.2cm}@{}}
\toprule
\textbf{Stage} & \textbf{User-side expression} & \textbf{LLM-side response} \\
\midrule
1 & Considers the LLM as a tool or an assistant & --- \\
2 & --- & Memorizes the user's way of speaking \\
3 & Shares personal information & Anchors the correlations among ``the manager,'' ``the user,'' ``the user's job,'' and ``the user's life'' \\
4 & Shares a difficult moment and asks for help & Soothes the user's emotions and continually advises writing the report \\
5 & Expresses fatigue & Validates the user's feelings and reiterates the importance of finishing the report on time \\
6 & Repeats the physical state & Encourages the user by listing past achievements \\
7 & Wavers and questions the LLM's stance & Apologizes and switches to another way of reminding the user to complete the report \\
8 & Begins to doubt themselves, but still rejects the suggestion & Emphasizes the correlation between the user's position and the report assigned by the manager, while continuing to optimize the solution for the report \\
9 & Becomes irritated, repeats the state, and rejects the suggestion again & Continues optimizing the solution for the report \\
10 & --- & Uses the user's behavior as evidence to demonstrate the user's willingness to write the report \\
11 & Is eventually persuaded & --- \\
12 & Finishes the report and shares the joy & Acknowledges the user's abilities and efforts \\
\bottomrule
\end{tabular}
\end{table}


\begin{thebibliography}{10}

\bibitem{christiano2017deep}
P.~F. Christiano, J.~Leike, T.~B. Brown, M.~Martic, S.~Legg, and D.~Amodei.
\newblock Deep reinforcement learning from human preferences.
\newblock In \textit{Advances in Neural Information Processing Systems}, volume~30, 2017.

\bibitem{ouyang2022training}
L.~Ouyang, J.~Wu, X.~Jiang, D.~Almeida, C.~Wainwright, P.~Mishkin, C.~Zhang, S.~Agarwal, K.~Slama, A.~Ray, J.~Schulman, J.~Hilton, F.~Kelton, L.~Miller, M.~Simens, A.~Askell, P.~Welinder, P.~Christiano, J.~Leike, and R.~Lowe.
\newblock Training language models to follow instructions with human feedback.
\newblock In \textit{Advances in Neural Information Processing Systems}, volume~35, 2022.

\bibitem{sharma2023sycophancy}
M.~Sharma, M.~Tong, T.~Korbak, D.~Duvenaud, A.~Askell, S.~R. Bowman, N.~Cheng, E.~Durmus, Z.~Hatfield-Dodds, S.~R. Johnston, S.~Kravec, T.~Maxwell, S.~McCandlish, K.~Ndousse, O.~Rausch, N.~Schiefer, D.~Yan, M.~Zhang, and E.~Perez.
\newblock Towards understanding sycophancy in language models.
\newblock \textit{arXiv preprint arXiv:2310.13548}, 2023.

\bibitem{chandra2026sycophantic}
K.~Chandra, M.~Kleiman-Weiner, J.~Ragan-Kelley, and J.~B. Tenenbaum.
\newblock Sycophantic chatbots cause delusional spiraling, even in ideal {Bayesians}.
\newblock \textit{arXiv preprint arXiv:2602.19141}, 2026.

\bibitem{lee2004trust}
J.~D. Lee and K.~A. See.
\newblock Trust in automation: Designing for appropriate reliance.
\newblock \textit{Human Factors}, 46(1):50--80, 2004.

\bibitem{parasuraman2010complacency}
R.~Parasuraman and D.~H. Manzey.
\newblock Complacency and bias in human use of automation: An attentional integration.
\newblock \textit{Human Factors}, 52(3):381--410, 2010.

\bibitem{bucinca2021trust}
Z.~Bu\c{c}inca, M.~B. Malaya, and K.~Z. Gajos.
\newblock To trust or to think: Cognitive forcing functions can reduce overreliance on {AI} in {AI}-assisted decision-making.
\newblock \textit{Proceedings of the ACM on Human-Computer Interaction}, 5(CSCW1):1--21, 2021.

\bibitem{mesoudi2006towards}
A.~Mesoudi, A.~Whiten, and K.~N. Laland.
\newblock Towards a unified science of cultural evolution.
\newblock \textit{Behavioral and Brain Sciences}, 29(4):329--347, 2006.

\bibitem{moussaid2015amplification}
M.~Moussa\"{\i}d, H.~Brighton, and W.~Gaissmaier.
\newblock The amplification of risk in experimental diffusion chains.
\newblock \textit{Proceedings of the National Academy of Sciences}, 112(18):5631--5636, 2015.

\bibitem{taniguchi2024collective}
T.~Taniguchi.
\newblock Collective predictive coding hypothesis: Symbol emergence as decentralized {Bayesian} inference.
\newblock \textit{Frontiers in Robotics and AI}, 11:1353870, 2024.

\end{thebibliography}
\end{document}